\documentclass[]{jkas} 

\def\year{2023} 
\def\volume{56} 
\def\issue{1} 
\def\beginpage{1} 
\def\received{May 26 2026} 
\def\accepted{July 10 2026} 
\def\published{---} 
\date{Received \received; Accepted \accepted; Published \published}

\newcommand\ion[2]{{#1}\,\textsc{#2}} 

\title{%
Origin of the Long-Period Radial Velocity Variation in the Red Supergiant HD 216946 (V424 Lac)
}

\author[1,2 $\star$]{Byeong-Cheol Lee}{0000-0002-2783-0755}
\author[3]{Myeong-Gu Park}{0000-0003-1544-8556}

\affil[1]{Korea Astronomy and Space Science Institute, 776, Daedeokdae-Ro, Youseong-Gu, Daejeon 34055, Korea}
\affil[2]{Korea University of Science and Technology, Gajeong-ro Yuseong-gu, Daejeon 34113, Korea}
\affil[3]{Department of Astronomy and Atmospheric Sciences, Kyungpook National University, Daegu 41566, Korea}

\def\corrauthor{%
Byeong-Cheol Lee, \email{bclee@kasi.re.kr}
}

\def\runningauthor{%
Byeong-Cheol Lee and Myeong-Gu Park
}

\def\runningtitle{%
Origin of the Long-Period Radial Velocity Variation in the Red Supergiant HD 216946 (V424 Lac)
}

\def\keywords{%
stars: individual: HD~216946 (V424 Lac) --- stars: supergiants ---
techniques: radial velocities --- stars: variables: general
}

\def\abstracttext{%
We present precise radial-velocity (RV) observations of the K-type red
supergiant HD~216946 (V424 Lac) obtained over approximately 22 years with the
Bohyunsan Optical Astronomy Observatory Echelle Spectrograph (BOES). The RV
measurements reveal a significant long-period variability with a period of
1365 days. To investigate its origin, we analyzed the RV data together with
line-profile variations (LPVs), chromospheric activity indicators, and
published photometric variability.
The LPVs exhibit periods of approximately 1370--1380 days, while the Na~D
lines show a similar periodicity near 1355 days, both comparable to the RV
period. In contrast, the H-line indicators display longer periods of
2730--2800 days, approximately twice the RV period. The close correspondence
between the RV variations and the activity-related diagnostics strongly
suggests that the 1365-day RV signal is primarily linked to chromospheric
activity and extended atmospheric variability rather than arising from purely
Keplerian motion. Published photometric studies also report additional
long-period variability, including a 1601-day long secondary-period
(LSP)-like variation. The coexistence of multiple non-identical
periods indicates that the observed variability of HD~216946 is unlikely to
originate from a single physical mechanism.
We therefore interpret HD~216946 as a multiperiodic red supergiant in which
several intrinsic stellar processes coexist. The observed variability is most
likely dominated by chromospheric activity and large-scale atmospheric
dynamics, possibly accompanied by rotational modulation and LSP-like
variability, although the presence of a low-mass stellar companion cannot be
completely excluded.
}

\begin{document}
\jkashead 


\section{Introduction\label{sec:intro}}
The radial velocity (RV) technique is one of the most successful methods for
detecting companions, providing orbital parameters and minimum companion
masses. Recent advances in high-precision spectroscopy have improved RV
precision to the level of $\sim1~\mathrm{m~s^{-1}}$ or better
\citep{Kambe2013}. While numerous companions have been discovered around
solar-mass main-sequence stars, detecting companions around evolved
intermediate-mass ($1.5$--$5\,M_{\odot}$) stars
\citep{Johnson2007,Bowler2010,Reffert2015} and evolved high-mass stars remains
challenging because these stars often exhibit substantial intrinsic RV
variability arising from pulsations, convection, atmospheric motions, and
stellar jitter \citep{Sato2005,Hekker2006,Josselin2007}.

Nevertheless, companion systems around intermediate- and high-mass stars
provide important constraints on theories of companion formation and orbital
evolution. Expanding the sample of such systems is therefore essential for
understanding how companion occurrence and properties depend on stellar mass
and evolutionary stage \citep{Hatzes2015,Lee2016}.

\begin{table*}[t]
\renewcommand{\thetable}{\arabic{table}}
\centering
\caption{Relative RV measurements for HD 216946 from 2004 to 2026 using the BOES.} \label{tab1}
\begin{tabular}{ccccccccc}
\hline
\hline
JD & RV  & $\pm \sigma$ & JD & RV  & $\pm \sigma$  & JD & RV  & $\pm \sigma$\\
$-$2,400,000 &{m\,s$^{-1}$}& {m\,s$^{-1}$}& $-$2,400,000 & {m\,s$^{-1}$} & {m\,s$^{-1}$} & $-$2,400,000 & {m\,s$^{-1}$} & {m\,s$^{-1}$}\\
\hline
53301.066719 &     163.5  &      6.1 &  56261.073198   &   112.5   &     9.9  &    58830.008117  &    376.0  &      9.4  \\
53301.073582 &     157.2  &      5.3 &  56331.930293   &   162.8   &     9.6  &    58859.918948  &    931.6  &     11.7  \\
53333.008976 &     281.8  &      5.2 &  56332.918772   &   157.4   &     9.6  &    58860.918480  &    975.5  &     11.9  \\
53899.285525 &     -90.7  &      6.7 &  56334.923775   &   185.4   &    12.1  &    58863.972649  &   1053.7  &     13.2  \\
54038.178477 &    -541.8  &      6.5 &  56336.907197   &   248.3   &     9.3  &    58933.346833  &    894.6  &     29.0  \\
54751.123932 &     515.7  &     10.5 &  56342.908530   &   383.0   &     8.8  &    59511.965259  &   -381.9  &     12.3  \\
54848.942883 &     663.9  &      7.2 &  56343.917202   &   418.8   &     9.6  &    59511.965259  &   -354.3  &     11.9  \\
54971.260309 &     562.7  &      7.5 &  56344.919301   &   442.4   &    10.9  &    59548.048960  &   -137.2  &     15.9  \\
54995.285493 &     369.1  &      6.5 &  56551.976596   &   286.4   &     9.2  &    59551.048800  &   -122.4  &     14.6  \\
55131.029603 &     139.7  &      6.4 &  56616.913973   &  -257.1   &     9.2  &    59561.923175  &   -100.8  &     13.3  \\
55356.235554 &   -1070.9  &      7.8 &  56616.916716   &  -255.8   &     9.2  &    59571.879158  &    -65.2  &     13.2  \\
55455.217130 &   -1113.8  &      7.5 &  56680.921648   &  -264.1   &    10.6  &    59718.212599  &   -501.1  &     20.6  \\
55553.982856 &   -1030.1  &      8.5 &  56680.923615   &  -268.6   &     8.9  &    59720.212688  &   -566.4  &     21.3  \\
55842.003562 &    -305.4  &      6.3 &  56680.925502   &  -263.0   &    10.2  &    59721.212733  &   -538.6  &     47.4  \\
55932.937119 &    -420.6  &      9.2 &  56790.293282   &  -621.5   &    29.6  &    61170.252189  &    109.9  &     18.9  \\
56177.044035 &     105.0  &      6.7 &  56790.298827   &  -572.4   &    23.4  &    61170.252189  &    148.7  &     19.1  \\

\hline
\end{tabular}
\end{table*}

Since 2003, we have conducted an RV survey to search for companions around
evolved stars. Within this program, the K-type supergiant HD~216946 was found
to exhibit low-amplitude, long-period RV variability \citep{Lee2014}.
Previous studies reported several long-term periodicities, including a
$\sim1350$-day RV signal, a photometric period of approximately
$\sim1100$ days detected in the \textit{HIPPARCOS} data, and a 1601-day
photometric variation identified from APT-80 observations
\citep{Messina2007}. 

HD~216946 is an important astrophysical target because it is a large and bright supergiant that may host a massive companion. As an evolved high-luminosity star, it provides an opportunity to study late-stage stellar evolution, including pulsation, atmospheric instability, and mass-loss processes. The possible presence of a massive companion also makes the system valuable for investigating long-period binary interactions and orbital evolution in evolved massive stars. Understanding HD~216946 may therefore provide important constraints on stellar evolution and massive binary systems.

\citet{Lee2014} suggested several possible interpretations for these
variations. The observed variability may be associated with semi-regular
variability in late-type supergiants, long secondary-period variability (LSP),
rotational modulation caused by surface inhomogeneities, or the presence of a
long-period companion. However, the available data at that time were
insufficient to determine the dominant mechanism conclusively.

The origin of the RV variability in HD~216946 therefore remains uncertain and
requires further investigation using additional diagnostics and improved
analysis techniques. In this study, we re-examine the RV variability of
HD~216946 using extended observational data together with activity-sensitive
and line-profile diagnostics. Section~2 describes the observations and data
reduction procedures. Section~3 presents the adopted stellar parameters.
Section~4 reports the orbital analysis of the RV data. Section~5 presents the
activity diagnostics and line-profile analyses, while Section~6 discusses the
possible origins of the observed RV variability. Finally, Section~7 summarizes
our conclusions.

\section{Observations and data reduction} \label{sec:style}
%
%
We obtained 48 high-resolution spectra of HD~216946 over approximately
22 years using the Bohyunsan Optical Astronomy Observatory Echelle Spectrograph
(BOES; \citealt{Kim2007}) mounted on the 1.8~m telescope at the Bohyunsan
Optical Astronomy Observatory (BOAO), Korea. BOES covers the wavelength range
3500--10500~\AA\ with a resolving power of $R \sim 90\,000$.

The observations were carried out between 2004 and 2026. The typical
signal-to-noise ratio in the iodine (I$_2$) region was about 150, and the
exposure times were limited to 15 minutes to minimize RV measurement errors.
The resulting RV precision for individual observations is approximately
$8~\mathrm{m~s^{-1}}$.

The spectra were reduced using the IRAF software package, and RVs were derived
from the iodine-cell calibrated spectra. The long-term RV stability of the
instrument was monitored using $\tau$~Ceti, for which a typical RV precision of
about $7~\mathrm{m~s^{-1}}$ was achieved \citep{Lee2013}. A summary of the
observations is given in Table~\ref{tab1}.

%
%


\begin{table}
\centering
\caption{Stellar parameters for HD~216946.}
\label{tab2}
\begin{tabular}{lcc}
\hline
Parameter & Value & Ref. \\
\hline
Spectral type & K5 Ib & 1 \\
$m_V$ (mag) & 4.99 & 2 \\
$B - V$ (mag) & $1.78 \pm 0.01$ & 2 \\
$\pi$ (mas) & $1.429 \pm 0.113$ & 3 \\
$T_{\rm eff}$ (K)  & $3767\pm 48$ & 4 \\
                   & $3940 \pm 30$ & 5 \\
$[$Fe/H$]$ & -- 0.039$^{+0.039}_{-0.035}$ & 4 \\
            & $-0.4 \pm 0.2$ & 5 \\
$\log g$ (cgs)  & $0.49 \pm 0.12$ & 4 \\
                & $0.8 \pm 0.3$ & 5 \\
$R_\star$ ($R_{\odot}$) & $265 \pm 22$ & 3 \\
$M_\star$ ($M_{\odot}$) & 9 -- 12  & 4 \\
$\log L_\star/L_\odot$ & 4.23 $\pm$ 0.10 & 4 \\
$v_{\rm micro}$ ($\mathrm{km~s^{-1}}$) & 1.98$^{+0.13}_{-0.11}$ & 4 \\
                                         & $2.3 \pm 0.2$ & 5 \\
$v_{\rm rot}\sin i$ ($\mathrm{km~s^{-1}}$) & $4.7 \pm 0.1$ & 4 \\
                                            & $4.6 \pm 0.2$ & 5 \\
\textbf{}$P_{\rm rot}/\sin i$ (days) & 2850 $\pm$ 250 & 5 \\
\hline
\end{tabular}

\vspace{2mm}
\noindent {\it References.}---(1) \citet{2000A&AS..143....9W}, (2) \citet{ESA1997}, (3) \citet{2023AA...674A...1G}, (4) \citet{2025A&A...693A.163T}, (5) This work
\end{table}


\section{\textbf{Stellar Characteristics}} \label{sec:Stellar properties}
HD~216946 is more consistent with a red supergiant (RSG) or an evolved massive supergiant than with a typical red giant branch (RGB) or low-mass asymptotic giant branch (AGB) star. Its high luminosity, large radius, low surface gravity, and luminosity class Ib support its classification as an evolved massive star.

The fundamental astrometric and kinematic parameters of HD~216946 were adopted from the \textit{HIPPARCOS} catalog \citep{ESA1997}. Stellar chemical abundances were compared with the results of \citet{2025A&A...693A.163T}, who developed and validated a near-infrared abundance analysis method using a sample of nearby RSGs.

We additionally derived atmospheric parameters from our high-resolution spectra using the TGVIT code of \citet{Takeda2005}, based on EW measurements of Fe~I and Fe~II lines. The projected rotational velocity ($v_{\rm rot}\sin i$) was estimated through line-profile fitting following the method of \citet{Takeda2008}. To distinguish rotational broadening from macroturbulent broadening, we adopted the empirical relation of \citet{Hekker2007},
\begin{equation}
v_{\rm macro} = 4.3 - 0.67 \log g \ \mathrm{km~s^{-1}} .
\end{equation}

The stellar chemical abundances derived in this study are in good agreement with those reported by \citet{2025A&A...693A.163T}. Adopting a stellar radius of $265 \pm 22\,R_\odot$ and a projected rotational velocity of $4.7\ \mathrm{km\ s^{-1}}$, we estimated a projected rotational period of
\[
P_{\rm rot}/\sin i = 2850 \pm 250~{\rm days}.
\]

The derived stellar parameters are summarized in Table~\ref{tab2}.


%
\section{Orbital solution}

%
\begin{table}
\centering
\caption{Tentative orbital solutions for HD 216946 derived from the MCMC simulation}
\label{tab3}
\begin{tabular}{lc}
\hline
Parameter & Value \\
\hline
P (days)              & 1365$^{+0.1}_{-0.1}$   \\
K (m s$^{-1}$)        & 722.9 $^{ +0.3}_{-0.2}$ \\
$e$                   & 0.13$^{+0.02}_{-0.02}$    \\
$T_{\rm periastron}$ (JD) & $2452718.4 \pm 106.3$ \\
Linear Slope  (m s$^{-1} d^{-1}$) & --0.18302157$^{+0.00186696}_{-0.00191305}$ \\
Quadratic Slope  (m s$^{-1} d^{-2}$) & --0.00004076$^{+0.00000043}_{-0.00000033}$  \\
RV offset (m s$^{-1}$) & -- 86.4$^{+0.3}_{-0.7}$ \\
RV jitter (m s$^{-1}$) &  2.8$^{+0.4}_{-0.2}$ \\
$N_{obs}$              & 48 \\
rms (m s$^{-1}$)       & 203  \\
\hline
\end{tabular}
\end{table}

\begin{figure}
\centering
\includegraphics[width=8.5cm]{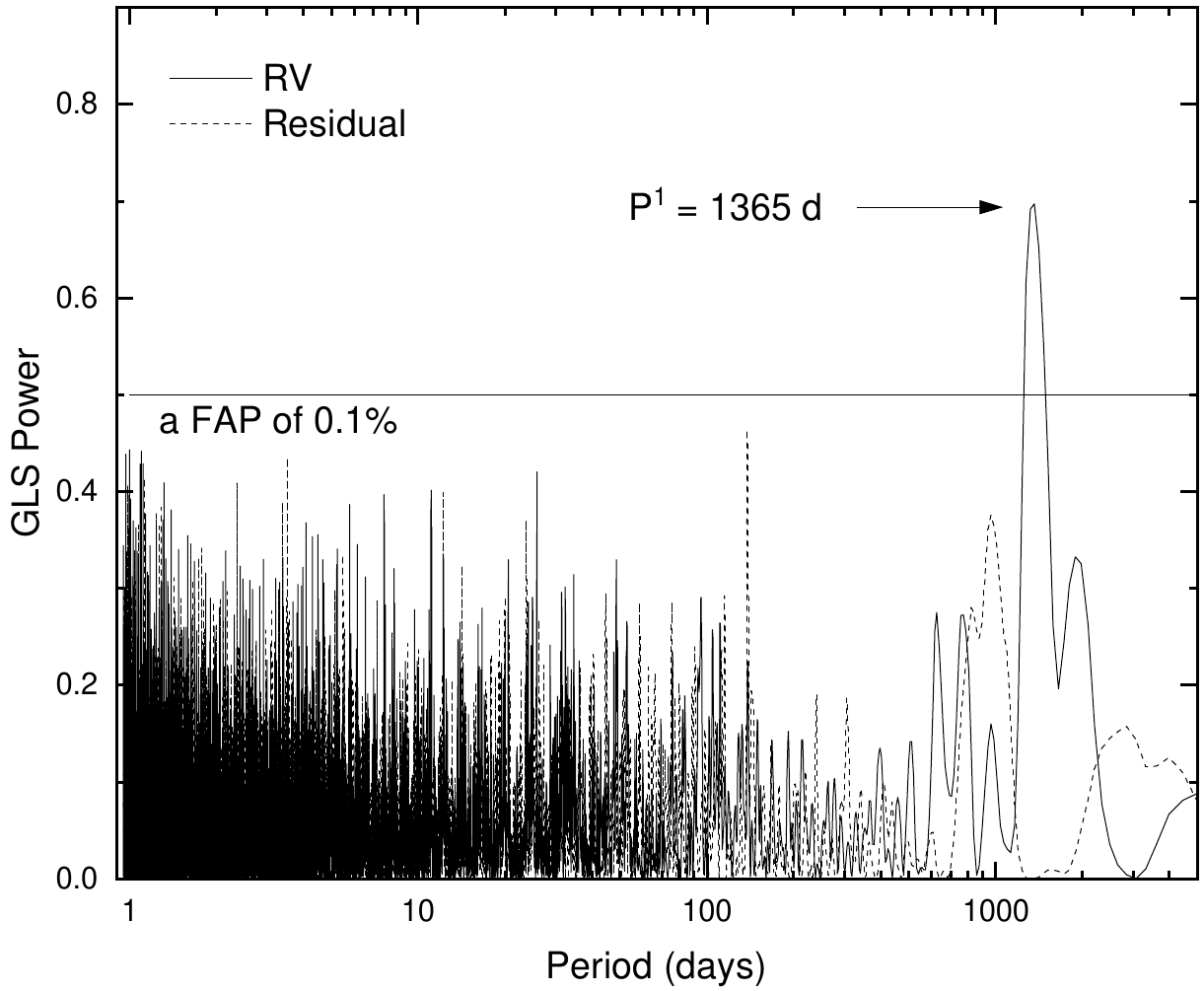}
\caption{The GLS periodograms for RV variations of HD~216946. The strongest peak occurs at 1365 days  and the periodogram for the residual RVs (dashed line) after subtracting the signal. The horizontal line  corresponds to a 0.1\% FAP. \label{f1}}
\end{figure}

\begin{figure}[t]
\centering
\includegraphics[width=8.5cm]{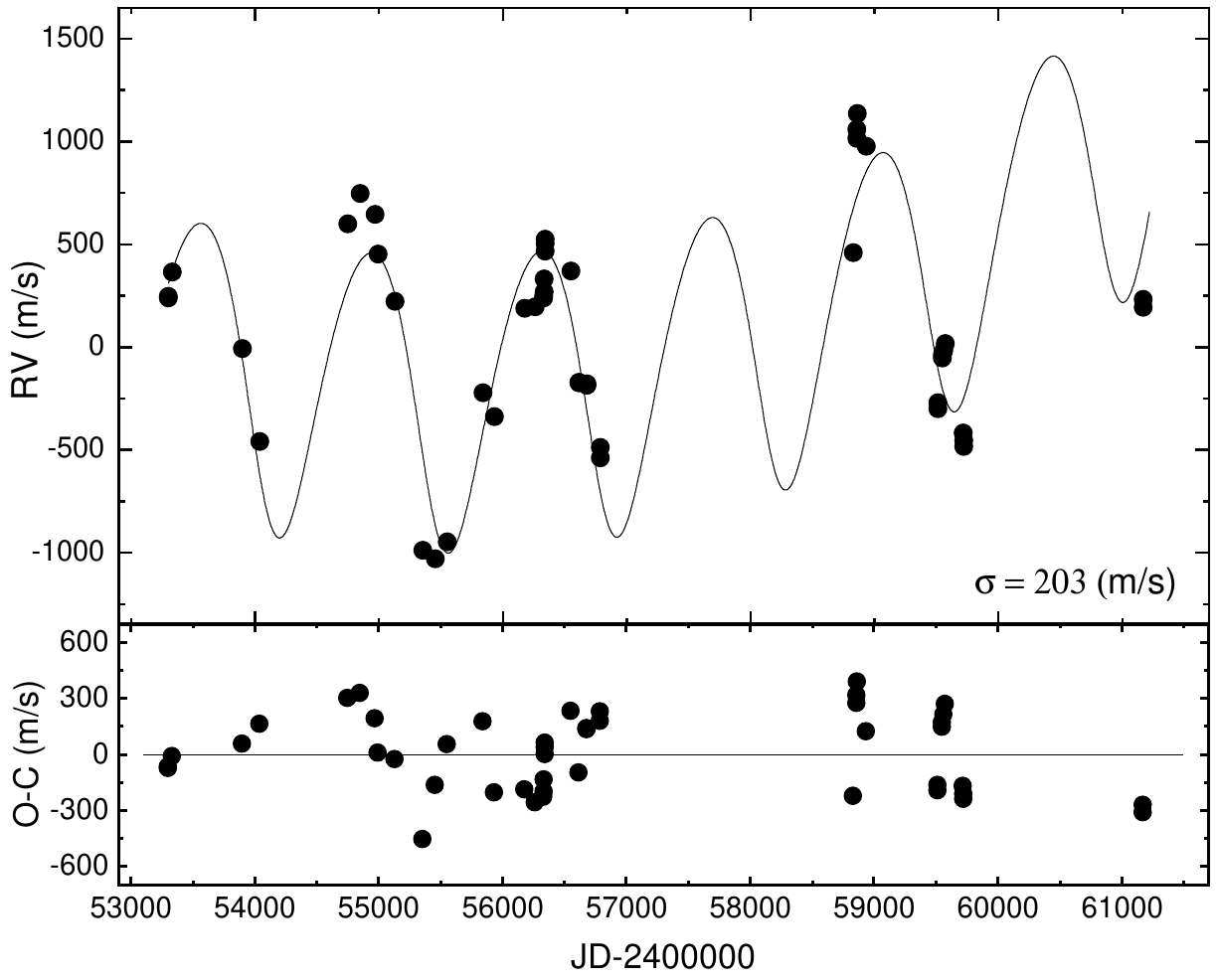}
\caption{The RV curve for HD~216946 is shown by the solid line.
The RV measurements and RV residuals are plotted in the upper and lower panels, respectively.
}
\label{f2}
\end{figure}

\begin{figure}[t]
\centering
   \includegraphics[width=8.5cm]{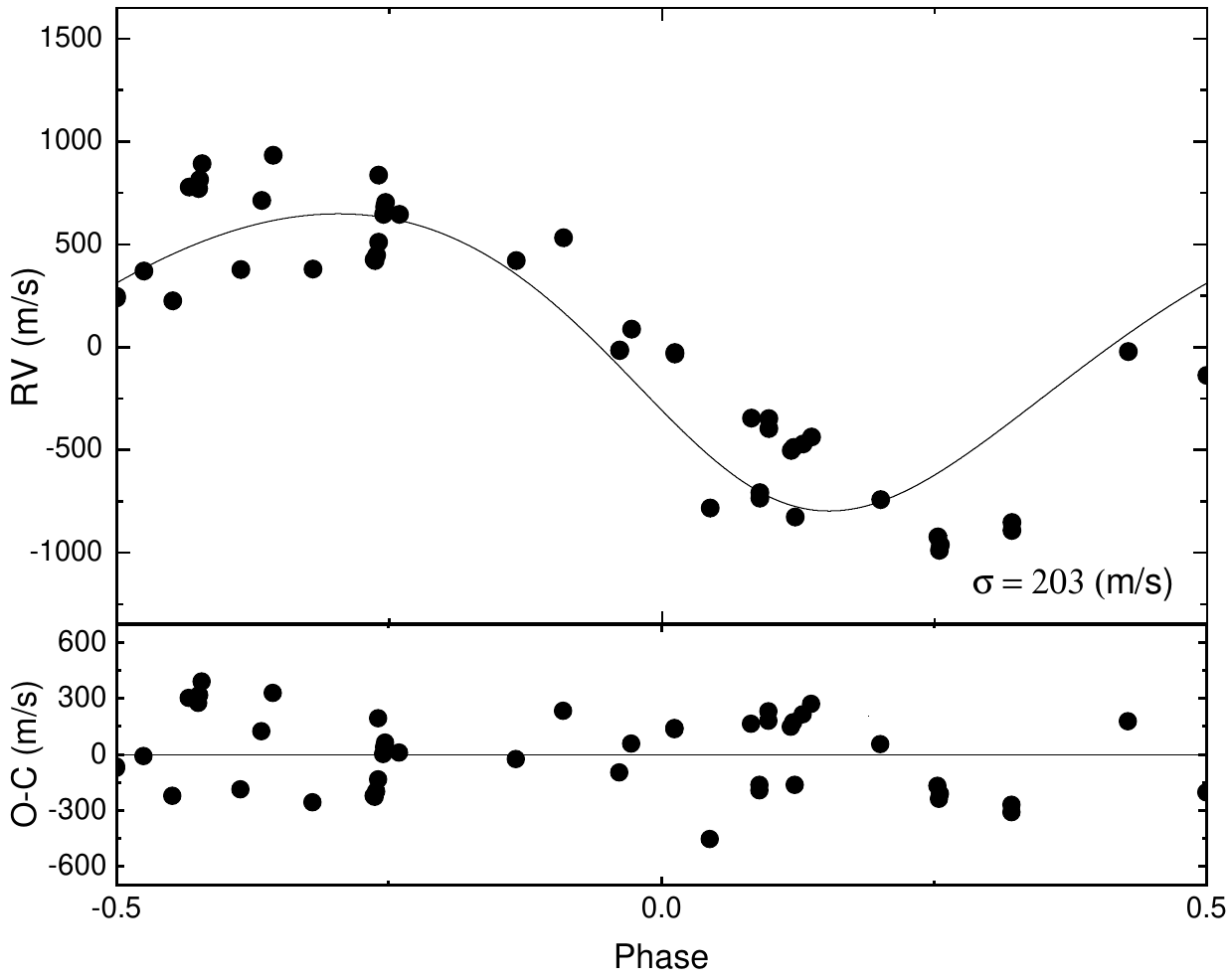}
\caption{The phase-folded RV curve with an orbital period of 1365~days  and  RV residuals are plotted in the upper and lower panels, respectively.
} \label{f3}
\end{figure}

We searched for periodic signals in the RV time series using the generalized
Lomb--Scargle periodogram (GLS; \citealt*{2009A&A...496..577Z}), which is
well suited for unevenly sampled data. The GLS periodogram reveals a significant
peak at approximately 1365 days (Figure~\ref{f1}). We confirmed that this
signal is not associated with the spectral window function or the observational
cadence.

To determine the tentative orbital parameters and their uncertainties, we performed a
Markov chain Monte Carlo (MCMC) analysis using the \texttt{Exo-Striker} tool
\citep{2019ascl.soft06004T}. The best-fit solution yields an orbital period of
$1365.0^{+0.1}_{-0.1}$ days and a semi-amplitude of
$722.9^{+0.3}_{-0.2}~\mathrm{m~s^{-1}}$. The resulting orbital parameters are
listed in Table~\ref{tab3}.

The RV time series of HD~216946 is shown in the upper panel of
Figure~\ref{f2}, and the corresponding phase-folded RV curve is presented in
Figure~\ref{f3}. After subtracting the 1365-day orbital solution and the linear
trend, the rms of the RV residuals is $203~\mathrm{m~s^{-1}}$, as shown in the
lower panel of Figure~\ref{f2}. No additional significant periodicity is
detected in the residuals.

Although the residual scatter is relatively large, such RV variability is not
unexpected for a luminous evolved star. \citet{Hekker2006} showed that RV
residuals in K giants tend to increase toward later spectral types and higher
luminosities, supporting the interpretation that the remaining scatter may be
partly associated with intrinsic stellar variability.


\begin{table*}[t]
\centering
\caption{Representative observational diagnostics associated with different mechanisms producing RV variability in evolved stars.}
\label{tab5}

\scriptsize
\renewcommand{\arraystretch}{1.03}
\setlength{\tabcolsep}{2pt}

\resizebox{\textwidth}{!}{
\begin{tabular}{lccccccc}
\hline\hline

Mechanism &
RV coherence &
Bisector behavior &
LPV characteristics &
Photometric variability &
H lines behavior &
Na\,I D lines \\

\hline

Radial pulsation       
& Moderate, periodic 
& Mild variation 
& Symmetric profiles 
& Strong periodicity 
& Mild variation 
& Mild variation \\

Non-radial pulsation   
& Moderate, multi-periodic 
& Large variation 
& Traveling bumps 
& Present 
& Possible variation 
& Possible variation \\

Rotational modulation  
& Moderately stable 
& BIS--RV correlation 
& Local asymmetry 
& Periodic modulation 
& Possible variation 
& Possible variation \\

Chromospheric activity 
& Moderate--low 
& Moderate variation 
& Core variability 
& Weak--moderate 
& Strong variation 
& Strong variation \\

Companion              
& Highly stable 
& Negligible 
& None 
& None 
& Stable 
& Stable \\

Convective variability             
& Low, stochastic 
& Large variation 
& Irregular profiles 
& Weak or quasi-periodic 
& Weak variation 
& Weak variation \\

Wind dynamics          
& Low, irregular 
& Complex variability 
& Variable core 
& Irregular variability 
& Strong variation 
& Moderate variation \\

\hline
\end{tabular}
}


\vspace{0.3mm}

\noindent {\it References.}---
\citep[e.g.,][]{Aigrain2012,Cincunegui2007,Cretignier2020,Gray1989,Hara2023,Hatzes2005,Hekker2006,Josselin2007,Kaufer1997,Kiss2006,Kravchenko2019,Lanza2011,Lyra2005,Meunier2019,Meunier2021,Montes1997,Saio2015,Xiong2007}.

\end{table*}


\section{Observational diagnostics}
Long-period RV variations in RSGs are unlikely to be
explained by a single physical mechanism. Owing to their extended atmospheres,
low surface gravities, vigorous convection, and substantial mass loss, RSGs can
exhibit RV variability produced by a combination of pulsation, convection,
atmospheric dynamics, rotational modulation, and, in some cases, orbital motion
due to a companion \citep[e.g.,][]{Kiss2006,Josselin2007,Kravchenko2019}.
Accordingly, a Keplerian interpretation of a long-period RV signal should be
regarded as tentative unless it is supported by long-term phase coherence and
by the absence of correlated spectroscopic or photometric activity indicators.

Table~\ref{tab5} summarizes key diagnostics for distinguishing companion-induced
RV signals from intrinsic stellar variability in evolved stars. Here, LPV denotes line-profile variation. Companion signals are expected to show long-term phase coherence with minimal LPV, whereas pulsation, convection, chromospheric activity, and atmospheric dynamics often produce correlated changes in bisector span, LPVs, and activity-sensitive features. Because multiple intrinsic mechanisms can coexist in cool supergiants and red supergiants, long-period RV signals should be assessed using RV
coherence, line-profile diagnostics, and activity indicators together.



\subsection{Bisector behavior}
Two bisector inverse slope (BIS) quantities, the bisector velocity span (BVS) and bisector
velocity curvature (BVC), were derived from the photospheric absorption line
Fe~I~6219.281\,\AA{}, which was selected for its strength, relative isolation,
and lack of significant blending or telluric contamination in the observed
spectra. Broad activity-sensitive features, saturated lines, and spectral
regions affected by atmospheric residuals were excluded to minimize spurious
profile distortions.

For each spectrum, the local continuum around Fe~I~6219.281\,\AA{} was
normalized, and the bisector was evaluated at the 40\% and 80\% flux levels
of the line depth. At each level, the wavelengths of the blue and red wings
were interpolated, and their midpoint was converted into a bisector velocity.
The BVS was defined as the velocity difference between the upper and lower
bisector levels, whereas the BVC quantified deviations from a linear bisector
shape, thereby tracing subtle line-profile asymmetries.

A GLS periodogram analysis was then applied to the resulting BVS and BVC time
series. No statistically significant periodicity was detected in either
diagnostic, as shown in Figure~\ref{f4}(a). 
The GLS false-alarm probability (FAP) was estimated using a
bootstrap randomization procedure \citep{2003A&A...403.1077K},
where the RV measurements were randomly shuffled
200,000 times while preserving the observing epochs.
The BVS periodogram shows a weak peak near
$\sim$1390~days with a FAP of approximately 3\%.
Although suggestive, this signal is not statistically
significant enough to support a correlation between the
line bisector variations and the observed RV variability.

\subsection{LPV characteristics }
To investigate the origin of the long-period RV variation in HD~216946, we
analyzed LPVs as diagnostics of photospheric velocity
fields and extended atmospheric dynamics. The LPV characteristics were examined
using the BVS, BVC, line depth, and EW.

The analysis was based on relatively strong and isolated photospheric metallic
lines in the 6000--6500~\AA\ region
\citep[e.g.,][]{HatzesCochran1998,HatzesCochran1999,HatzesEtAl2006}. Among
them, Fe~I 6265.132~\AA\ was adopted as the primary LPV diagnostic because it
is strong and minimally affected by blends or telluric contamination.
GLS periodogram analyses of the Fe~I 6265.132~\AA\ line revealed significant
periodicities of approximately 1370 and 1380~days in the BVC and BVS,
respectively [Figure~\ref{f4}(b)].

In addition, the independent diagnostic
line Ca~I 6439.08~\AA\ showed a consistent period of approximately 1371~days
in both the line depth and EW measurements [Figure~\ref{f4}(c)]. 

These 1370--1380~day periodicities are consistent with the 1365-day RV period,
suggesting that the RV variability is closely related to changes in the
photospheric velocity field or extended line-forming regions.

\subsection{Photometric variability}
The \textit{HIPPARCOS} photometric data for HD~216946 were obtained between
December 1989 and February 1993, providing a total of 173 individual
measurements. These observations offer an independent constraint on possible
long-term brightness variations associated with the RV signal. The
\textit{HIPPARCOS} light curve shows a relatively small photometric scatter,
with an rms below 0.013~mag, indicating that the star was photometrically
stable at the millimagnitude level during the observing interval.

To search for periodic brightness variations, we applied a GLS periodogram
analysis to the \textit{HIPPARCOS} photometric time series. The resulting
periodogram is shown in Figure~\ref{f4}(g). A prominent peak is detected at
approximately 1098~days.

\subsection{H \& D lines behaviors} 
The strengths of \ion{Ca}{ii} H \& K, H$_{\alpha}$, H$_{\beta}$,
\ion{Mg}{i}~b, \ion{He}{i}~D$_3$, \ion{Na}~D$_1$ and D$_2$, and the
\ion{Ca}{ii} infrared triplet (IRT) lines are widely used as diagnostics of
chromospheric activity because they probe different atmospheric layers.
However, in the present BOES spectra of HD~216946, the \ion{Ca}{ii} H \& K
region has insufficient signal-to-noise ratio, the \ion{Ca}{ii} IRT lines are
affected by CCD saturation beyond $\sim6800$~\AA, and the \ion{Mg}{i}~b
triplet lies within the I$_2$ absorption region. We therefore focused on the
H and Na~D lines as the primary diagnostics of chromospheric and atmospheric
variability.

\begin{figure}
   \centering
   \includegraphics[width=8.5cm]{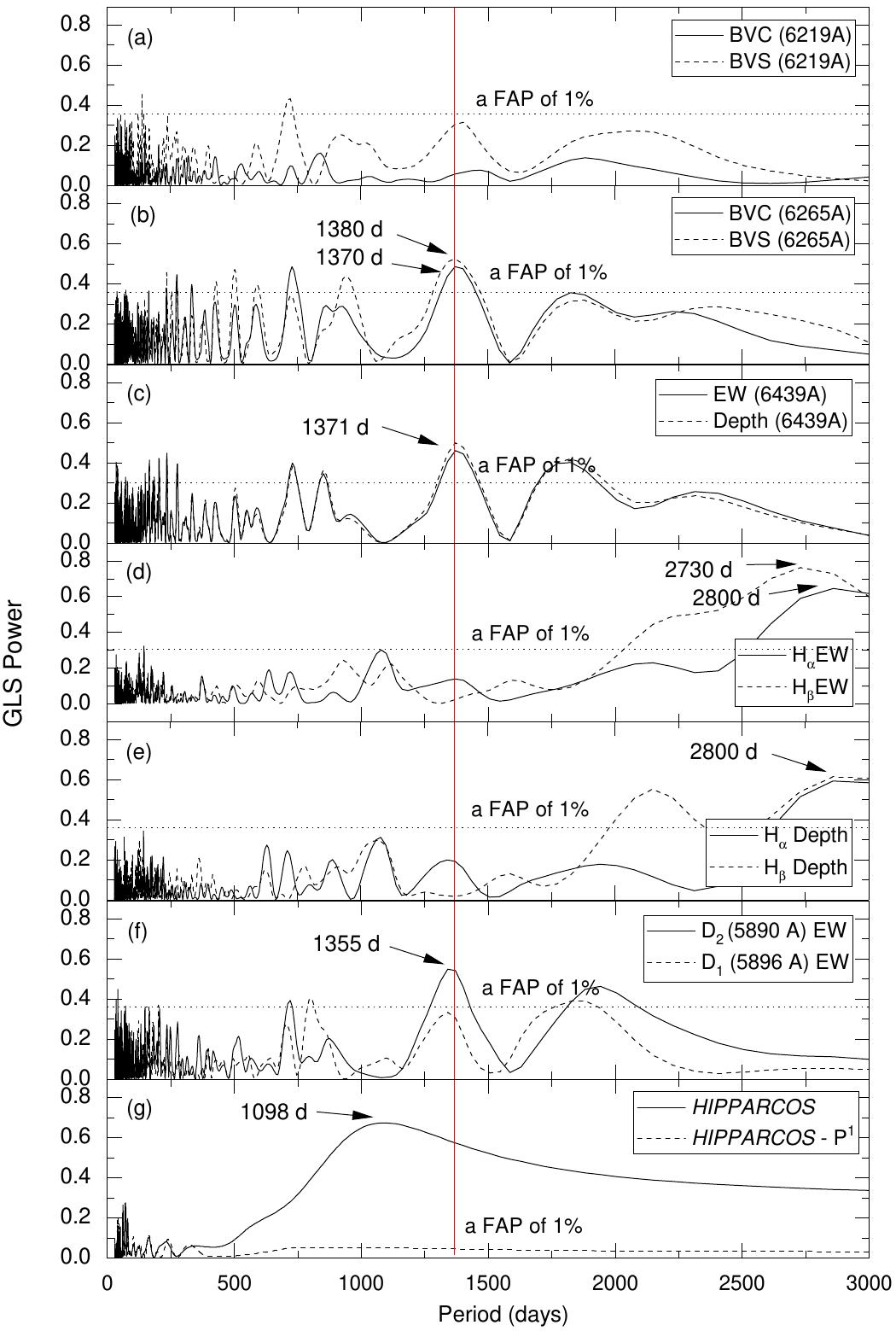}
      \caption{GLS periodograms of various diagnostic indicators for HD~216946.
Panels show the periodograms of (a) the line-bisector indicators of the
reference photospheric line, (b) the LPV bisector indicators, (c) the EW and
line depth of the LPV diagnostic lines, (d) the H line bisector indicators,
(e) the H-line EW and depth indicators, (f) the Sodium line bisector indicators,
and (g) the \textit{HIPPARCOS} photometric data. The horizontal dashed line
in each panel indicates the 1\% FAP level and 
the vertical red solid line marks the RV period of 1365~days.}
        \label{f4}
   \end{figure}

The H$_{\alpha}$ line is particularly useful for late-type stars because of
their increasing flux toward red wavelengths
\citep{Cincunegui2007}. In addition, the relatively weak telluric
contamination and narrow hydrogen absorption cores allow reliable EW
measurements. For HD~216946, the EWs of the H$_{\alpha}$ and H$_{\beta}$ lines
were measured using passbands of $\pm1.0$~\AA\ and $\pm0.8$~\AA,
respectively, centered on each line core to minimize contamination from nearby
blends and atmospheric H$_2$O absorption. The GLS periodograms of the H-line
EWs and line depths show periods of approximately 2730--2800~days
[Figure~\ref{f4}(d,e)], close to twice the 1365-day RV period.

The Na~D$_2$ and Na~D$_1$ EWs were measured at 5889.951~\AA\ and
5895.924~\AA, respectively, using a $\pm0.5$~\AA\ bandpass centered on each
line core. The Na~D lines exhibit a period of approximately 1355~days
[Figure~\ref{f4}(f)], comparable to the 1365-day RV period.

%


\section{Possible Origins of the RV variations}

Long-period RV variability in K-type red supergiants may arise from several
mechanisms, including pulsation, convection, rotational modulation,
atmospheric dynamics, mass loss, and orbital motion due to a companion.
Pulsations and large convective cells can produce photometric and line-profile
variations together with spectral-line asymmetries and apparent RV shifts that
may mimic companion-induced signals
\citep{Kiss2006,Kravchenko2019}. LSPs, commonly
observed in luminous red variables, may also contribute to long-period
variability, although their physical origin remains uncertain
\citep{Kiss2006}. In extended RSG atmospheres, atmospheric motions, wind
variability, and episodic mass ejection can introduce depth-dependent RV
shifts, and the measured RV may depend on the spectral lines used because
different atmospheric layers can exhibit different velocity fields
\citep{Josselin2007,Kravchenko2019}. In contrast, a genuine companion is
expected to produce a stable and coherent RV signal with little or no
correlated photometric, chromospheric, or line-profile variability.
Therefore, reliable interpretation of long-period RV signals requires combined
analyses of RV coherence, LPVs, activity-sensitive lines, and photometric
variability.

\subsection{Pulsation}
Using the revised oscillation-amplitude scaling relation of
\citet{2011A&A...529L...8K}, we estimated a pulsation-induced RV amplitude of
$310^{+90}_{-70}~\mathrm{m~s^{-1}}$ for HD~216946. The corresponding radial
pulsation period inferred from the $\nu_{\max}$ scaling relation is
$20.3^{+4.0}_{-3.5}$ days, while the fundamental dynamical period is
$154^{+21}_{-20}$ days. These predicted pulsation timescales and amplitudes are
significantly smaller than the observed RV signal, which has a semi-amplitude
of approximately $723~\mathrm{m~s^{-1}}$ and a period of 1365 days. Therefore,
standard solar-like radial pulsations alone are unlikely to explain the
dominant long-period RV variability, although intrinsic stellar variability may
still contribute to the residual RV scatter.

HD~216946 was monitored for approximately eight years (August 1993 to October
2001) using the 80-cm automated photometric telescope (APT-80) at the
INAF-Catania Astrophysical Observatory \citep{Messina2007}. That study reported
clear photometric variability with amplitudes of approximately
$0.08$~mag in the $V$ band, $0.11$~mag in the $B$ band, and $0.33$~mag in the
$U$ band, substantially larger than the rms scatter of the \textit{HIPPARCOS}
photometry. Several shorter periodicities ($P = 305$, 422, and 490~days), as
well as a longer-term variation with a period of $1601$~days, were identified.

In our analysis of the \textit{HIPPARCOS} photometry, we detected a periodicity
near 1098 days, suggesting the presence of additional long-term intrinsic
variability. However, this period differs from both the 1365-day RV period
identified in this study and the 1601-day LSP-like
variation reported by \citet{Messina2007}. Consequently, the origin of the
1098-day signal remains uncertain, and the currently available photometric and
spectroscopic diagnostics are insufficient to establish a clear pulsational
interpretation.


\subsection{Rotational modulation}
Rotational modulation is a possible source of long-period RV variability in
evolved stars. Large surface inhomogeneities, such as giant spots, temperature
structures, magnetic active regions, or convective features, can distort
spectral-line profiles as the star rotates and thereby induce apparent RV
shifts \citep{2001A&A...379..279Q,Hatzes2005,Lanza2011,Meunier2019}. Such
variability is generally expected to be accompanied by photometric modulation,
line-profile asymmetries, and bisector variations. In cool supergiants and red
supergiants, however, rotational modulation is often difficult to distinguish
from convection-driven variability because both processes can produce
quasi-periodic RV signals and LPVs.

A GLS analysis of the \textit{HIPPARCOS} photometry of HD~216946
reveals a broad excess of power at timescales longer than
approximately 1100\,days. Because this characteristic
timescale is comparable to the observational baseline, the
available data do not permit a reliable determination of a
unique period, and the associated FAP is correspondingly uncertain. 
We therefore interpret the \textit{HIPPARCOS} photometry only as evidence for
possible long-timescale variability, without assigning a
specific periodicity or physical origin. No corresponding
periodicity is detected in the bisector diagnostics, and
therefore the available data do not support a direct link
between the long-timescale photometric variability and
line-profile variations, rotational modulation, or stellar
pulsations.

On the other hand, we estimated the projected rotational period,
$P_{\rm rot}/\sin i = 2850 \pm 250~{\rm days}$, using the stellar radius and
projected rotational velocity. If the stellar spin axis is viewed nearly
edge-on, i.e., $\sin i \simeq 1$, the true rotational period would be close to
this value. This timescale is comparable to the $\sim2730$--$2800~{\rm days}$
periodicity detected in the EWs and line depths. Therefore, rotational
modulation cannot be excluded as a possible contributor, at least, partly  
to the observed spectroscopic variability.

\subsection{Chromospheric activity}
Chromospheric activity diagnostics were used to investigate magnetic and
atmospheric variability in HD~216946. The H lines, including H$\alpha$ and
H$\beta$, primarily trace activity in the middle chromosphere, whereas the
Na~I D lines probe the upper photosphere and lower chromosphere. We therefore
adopted the H and Na~D lines as the principal diagnostics of chromospheric
activity and extended atmospheric variability.

In addition, LPV diagnostics based on the photospheric lines Fe~I
6265.132~\AA\ and Ca~I 6439.08~\AA\ revealed significant periodicities of
approximately 1370--1380~days in the BVS, BVC, line depth, and EW
measurements [Figure~\ref{f4}(b,c)]. These periods are consistent with the
1365-day RV period, indicating that the RV variability is closely linked to
changes in the photospheric velocity field or extended line-forming regions.

The Na~D lines exhibit a $\sim1355$-day period comparable to the RV period,
whereas the H-line equivalent widths and line depths show longer periods of
approximately 2730--2800~days. The agreement between the RV period and the
periodicities detected in the Na~D and LPV diagnostics, together with the
longer H-line variability, suggests that the 1365-day RV signal is more likely
associated with chromospheric activity and/or extended atmospheric dynamics
than produced solely by Keplerian motion.

\subsection{Companion}
The observed periodic RV variations may well be described by a coherent
Keplerian model, suggesting the possible presence of an orbiting companion.
The best-fit orbital solution yields a minimum companion mass of
$M_{\rm c}\sin i = 170$--$208\,M_{\mathrm{J}}$, an orbital period of
$1365^{+0.1}_{-0.1}$ days, and a semi-major axis of approximately
5.0--5.5 AU.

\citet{2022A&A...657A...7K} used astrometric data from both \textit{Gaia} DR2
and \textit{HIPPARCOS} to investigate long-term stellar accelerations. For
HD~216946, assuming a stellar mass of $6.5\,M_\odot$ and a characteristic
inclination of $60^{+21}_{-27}$ degrees, they inferred a possible companion
with a mass of $229^{+121}_{-100}\,M_{\mathrm{J}}$ at a projected separation
normalized to 5 AU.

Although the masses inferred from the RV and astrometric analyses are not
identical, both methods point to a massive companion in the low-mass stellar
regime. The difference can plausibly be explained by the different adopted
stellar masses, inclination assumptions, projected separation, and orbital
geometry. In particular, the RV solution provides only a minimum mass, whereas
the astrometric estimate is more directly sensitive to the true companion mass.
Thus, the two results are broadly compatible within the systematic
uncertainties of the methods.

Nevertheless, the companion interpretation should be regarded as tentative.
Luminous evolved supergiants can exhibit long-period RV variability driven by
intrinsic stellar processes such as pulsations, large-scale convection,
chromospheric activity, and atmospheric dynamics. Therefore, although the
presence of a companion cannot be excluded, the Keplerian interpretation
requires further confirmation through additional observational diagnostics.
If confirmed, this system would represent a noteworthy case, since companions
appear to be relatively uncommon around intermediate-mass giant stars
\citep{Reffert2015} and are even more rarely identified around supergiants.

\subsection{Convection}

Convection is considered one of the primary intrinsic sources of RV variability
in cool supergiants and red supergiants. Owing to their extended envelopes and
low surface gravities, large convective cells can generate substantial
photospheric velocity fields, producing spectral-line asymmetries, bisector
variations, and apparent RV shifts \citep{Gray1989,Josselin2007,Kravchenko2019}.
Convection-driven RV variability is generally quasi-periodic or stochastic in
nature and is often accompanied by LPVs and limited
phase coherence, which can mimic RV signals induced by orbiting companions.

In the present study, however, we found no clear spectroscopic diagnostics that
strongly support convection as the dominant origin of the observed long-period
RV variability.

\subsection{Wind dynamics}
Wind dynamics and atmospheric motions are considered important sources of RV
variability in supergiants, particularly in red supergiants with extended
atmospheres and substantial mass loss. Variations in stellar winds,
atmospheric circulation, shock propagation, and episodic mass ejection can
produce asymmetric line profiles, variable line cores, and apparent RV shifts
\citep{Josselin2007,Gray2008,Harper2009}. These wind-driven signals are
typically irregular or weakly coherent and often accompanied by strong
variability in activity-sensitive lines such as H$\alpha$. Considering the
H-line analysis performed in this study, the influence of stellar winds and
atmospheric motions cannot be neglected as possible contributors to the
observed RV variability.

\section{Discussion\label{sec:dis}}
We investigated the origin of the 1365-day RV periodicity detected in the red
supergiant HD~216946 (V424 Lac) using RV measurements, photometry,
LPVs, and chromospheric activity indicators. The
LPVs show periods of approximately 1370--1380~days, while the Na~D lines
exhibit a similar periodicity near 1340~days. In contrast, the H-line
indicators display a longer timescale close to twice the RV period. These
results suggest that the 1365-day RV signal is more likely associated with
atmospheric and chromospheric variability than with purely Keplerian motion,
consistent with previous studies of intrinsic RV variability in red
supergiants \citep[e.g.,][]{Kiss2006,Josselin2007,Chiavassa2011,Patrick2019,Ahmad2023,Lee2023}.

HD~216946 also exhibits complex long-timescale photometric variability.
The 1601-day period reported by \citet{Messina2007} has been interpreted
as an LSP-like variation, whereas the \textit{HIPPARCOS} photometry
suggests only variability on timescales longer than approximately
1100 days, since the available observational baseline is insufficient
to constrain a unique period. The coexistence of multiple non-identical
periods suggests that the observed variability does not arise from a single
physical mechanism. Similar multiperiodic behavior in luminous red giants and
supergiants has been linked to pulsations, oscillatory convective modes,
binarity, dust-related effects, and large-scale atmospheric dynamics
\citep[e.g.,][]{Wood1999,Kiss2006,Saio2015,Trabucchi2021}.

 The 1365-day RV signal is more plausibly
associated with activity-related processes in the extended
stellar atmosphere, as supported by the corresponding
periodicities detected in the chromospheric activity indicators.

Overall, HD~216946 is most plausibly interpreted as a multiperiodic red
supergiant in which several intrinsic processes coexist.
The \textit{HIPPARCOS} photometry suggests the presence of
photometric variability on timescales longer than
approximately 1100\,d, although the available observations
are insufficient to constrain a unique period or identify
its physical origin. In contrast, the $\sim$1601-day
photometric variation reported by \citet{Messina2007} has
been interpreted as LSP-like variability or long-period pulsation.
The 1365-day RV signal is more likely linked to chromospheric
activity and extended atmospheric dynamics. Although the presence of a
low-mass companion cannot be completely excluded, the correspondence between
the RV period and the activity-related diagnostics favors an intrinsic stellar
origin for the observed long-period RV variability.


\acknowledgments
BCL acknowledges partial support by the KASI (Korea Astronomy and Space Science Institute) grant
2026-1-904-01.
MGP was supported by the Basic Science Research Program through the NRF funded by the Ministry of Education (RS-2019-NR045193, RS-2018-NR031074) under the R\&D program supervised by the Ministry of Science, ICT and Future Planning.





\end{document}